\documentclass[aps,english,twocolumn,pra]{revtex4-1}
\usepackage{amsmath}
\usepackage{SIunits}
\usepackage{graphicx}% Include figure files
\begin{document}
%
%
% Define new commands:
%
\newcommand{\ac}[0]{\ensuremath{\hat{a}_{\mathrm{c}}}}
\newcommand{\adagc}[0]{\ensuremath{\hat{a}^{\dagger}_{\mathrm{c}}}}
\newcommand{\aR}[0]{\ensuremath{\hat{a}_{\mathrm{R}}}}
\newcommand{\aT}[0]{\ensuremath{\hat{a}_{\mathrm{T}}}}
\renewcommand{\b}[0]{\ensuremath{\hat{b}}}
\newcommand{\bdag}[0]{\ensuremath{\hat{b}^{\dagger}}}
\newcommand{\betaI}[0]{\ensuremath{\beta_\mathrm{I}}}
\newcommand{\betaR}[0]{\ensuremath{\beta_\mathrm{R}}}
\renewcommand{\c}[0]{\ensuremath{\hat{c}}}
\newcommand{\cdag}[0]{\ensuremath{\hat{c}^{\dagger}}}
\newcommand{\CorrMat}[0]{\ensuremath{\boldsymbol\gamma}}
\newcommand{\Deltacs}[0]{\ensuremath{\Delta_{\mathrm{cs}}}}
\newcommand{\Deltacsmax}[0]{\ensuremath{\Delta_{\mathrm{cs}}^{\mathrm{max}}}}
\newcommand{\Deltacsparked}[0]{\ensuremath{\Delta_{\mathrm{cs}}^{\mathrm{p}}}}
\newcommand{\Deltacstarget}[0]{\ensuremath{\Delta_{\mathrm{cs}}^{\mathrm{t}}}}
\newcommand{\Deltae}[0]{\ensuremath{\Delta_{\mathrm{e}}}}
\newcommand{\Deltahfs}[0]{\ensuremath{\Delta_{\mathrm{hfs}}}}
\newcommand{\dens}[0]{\ensuremath{\hat{\rho}}}
\newcommand{\erfc}[0]{\ensuremath{\mathrm{erfc}}}
\newcommand{\Fq}[0]{\ensuremath{F_{\mathrm{q}}}}
\newcommand{\gammapar}[0]{\ensuremath{\gamma_{\parallel}}}
\newcommand{\gammaperp}[0]{\ensuremath{\gamma_{\perp}}}
\newcommand{\gavg}[0]{\ensuremath{\mathcal{G}_{\mathrm{avg}}}}
\newcommand{\gbar}[0]{\ensuremath{\bar{g}}}
\newcommand{\gens}[0]{\ensuremath{g_{\mathrm{ens}}}}
\renewcommand{\H}[0]{\ensuremath{\hat{H}}}
\renewcommand{\Im}[0]{\ensuremath{\mathrm{Im}}}
\newcommand{\kappac}[0]{\ensuremath{\kappa_{\mathrm{c}}}}
\newcommand{\kappamin}[0]{\ensuremath{\kappa_{\mathrm{min}}}}
\newcommand{\kappamax}[0]{\ensuremath{\kappa_{\mathrm{max}}}}
\newcommand{\ket}[1]{\ensuremath{|#1\rangle}}
\newcommand{\mat}[1]{\ensuremath{\mathbf{#1}}}
\newcommand{\mean}[1]{\ensuremath{\langle#1\rangle}}
\newcommand{\omegac}[0]{\ensuremath{\omega_{\mathrm{c}}}}
\newcommand{\omegas}[0]{\ensuremath{\omega_{\mathrm{s}}}}
\newcommand{\pauli}[0]{\ensuremath{\hat{\sigma}}}
\newcommand{\pexc}[0]{\ensuremath{p_{\mathrm{exc}}}}
\newcommand{\pexceff}[0]{\ensuremath{p_{\mathrm{exc}}^{\mathrm{eff}}}}
\newcommand{\Pa}[0]{\ensuremath{\hat{P}_{\mathrm{c}}}}
\newcommand{\Qmin}[0]{\ensuremath{Q_{\mathrm{min}}}}
\newcommand{\Qmax}[0]{\ensuremath{Q_{\mathrm{max}}}}
\renewcommand{\Re}[0]{\ensuremath{\mathrm{Re}}}
\renewcommand{\S}[0]{\ensuremath{\hat{S}}}
\newcommand{\Sminuseff}[0]{\ensuremath{\hat{S}_-^{\mathrm{eff}}}}
\newcommand{\Sxeff}[0]{\ensuremath{\hat{S}_x^{\mathrm{eff}}}}
\newcommand{\Syeff}[0]{\ensuremath{\hat{S}_y^{\mathrm{eff}}}}
\newcommand{\tildeac}[0]{\ensuremath{\tilde{a}_{\mathrm{c}}}}
\newcommand{\tildepauli}[0]{\ensuremath{\tilde{\sigma}}}
\newcommand{\Tcaveff}[0]{\ensuremath{T_{\mathrm{cav}}^{\mathrm{eff}}}}
\newcommand{\Techo}[0]{\ensuremath{T_{\mathrm{echo}}}}
\newcommand{\Tmem}[0]{\ensuremath{T_{\mathrm{mem}}}}
\newcommand{\Tswap}[0]{\ensuremath{T_{\mathrm{swap}}}}
\newcommand{\Var}[0]{\ensuremath{\mathrm{Var}}}
\renewcommand{\vec}[1]{\ensuremath{\mathbf{#1}}}
\newcommand{\Xa}[0]{\ensuremath{\hat{X}_{\mathrm{c}}}}

\title{Storage and retrieval of microwave fields at the single-photon level in a spin ensemble}

\author{C. Grezes$^{1}$, B. Julsgaard$^{2}$, Y. Kubo$^{1}$, W. L. Ma$^{3,4}$, M. Stern$^{1}$, A. Bienfait$^{1}$, K. Nakamura$^{5}$, J. Isoya$^{6}$, S. Onoda$^{7}$, T. Ohshima$^{7}$, V. Jacques$^{8}$, D. Vion$^{1}$, D. Esteve$^{1}$, R. B. Liu$^{3}$, K. M{\o}lmer$^{2}$, and P. Bertet$^{1}$}

\affiliation{$^{1}$Quantronics group, Service de Physique de l'Etat Condensé, DSM/IRAMIS/SPEC, CNRS UMR 3680, CEA Saclay,
91191 Gif-sur-Yvette, France}

\affiliation{$^{2}$Department of Physics and Astronomy, Aarhus University, Ny Munkegade 120, DK-8000 Aarhus C, Denmark.}

\affiliation{$^{3}$Department of Physics, Centre for Quantum Coherence, and Institute of Theoretical Physics, The Chinese University of Hong Kong, Shatin, New Territories, Hong Kong, China}

\affiliation{$^{4}$State Key Laboratory of Superlattices and Microstructures, Institute of Semiconductors, Chinese Academy of Sciences, Beijing 100083, China}
	
\affiliation{$^{5}$Energy System Research Institute, Fundamental Technology Department, Tokyo Gas Co., Ltd., Yokohama, 230-0045, Japan}

\affiliation{$^{6}$Research Center for Knowledge Communities, University of
Tsukuba, Tsukuba 305-8550, Japan}

\affiliation{$^{7}$Japan Atomic Energy Agency, Takasaki 370-1292, Japan}

\affiliation{$^{8}$Laboratoire Aim{\'e} Cotton, CNRS, Université Paris-Sud and ENS Cachan, 91405 Orsay, France}

%Collaboration name if desired (requires use of superscriptaddress
%option in \documentclass). \noaffiliation is required (may also be
%used with the \author command).
%\collaboration can be followed by \email, \homepage, \thanks as well.
%\collaboration{}
%\noaffiliation

\date{\today}

\begin{abstract}
We report the storage of microwave pulses at the single-photon level in a spin-ensemble memory consisting of $10^{10}$ NV centers in a diamond crystal coupled to a superconducting LC resonator. The energy of the signal, retrieved $100\, \mu \mathrm{s}$ later by spin-echo techniques, reaches $0.3\%$ of the energy absorbed by the spins, and this storage efficiency is quantitatively accounted for by simulations. This figure of merit is sufficient to envision first implementations of a quantum memory for superconducting qubits.
\vspace{1cm}
\end{abstract}
\maketitle

Superconducting qubits are attractive candidates for solid-state implementations of quantum information processing, but suffer from coherence times shorter than $\sim 100 \mu \mathrm{s}$~\cite{Rigetti.PhysRevB.86.100506(2012),Barends.PhysRevLett.111.080502(2013),Stern.PhysRevLett.113.123601(2014)}. To circumvent this issue, it has been proposed to use ensembles of spins in semiconductors\cite{Kubo.PhysRevLett.105.140502(2010),Schuster.PhysRevLett.105.140501(2010),Amsuss.PhysRevLett.107.060502(2011),Ranjan.PhysRevLett.110.067004,Probst.PhysRevLett.110.157001(2013),Tabuchi.PhysRevLett.113.083603(2014)} as a multimode quantum memory, able to store multiple qubit states over longer periods of time, and to retrieve them on-demand~\cite{Kurizki.PNASquantum(2015)}. Inspired by research on optical quantum memories~\cite{Damon.NewJPhys.13.093031(2011),McAuslan.PhysRevA.84.022309(2011),Rielander.PhysRevLett.112.040504(2014)}, realistic protocols have been proposed recently~\cite{Julsgaard.PhysRevLett.110.250503,Afzelius.NJP.1367-2630-15-6-065008(2013)}. The state of a superconducting qubit is first converted into the state of a microwave photon, propagating or trapped in a resonator. This photon is then resonantly and collectively absorbed by the spin ensemble, resulting in a transverse magnetization which, due to the spread of resonance frequency within the ensemble, decays in a time $T_2^*$ called the free-induction decay (FID) time. The {\it write} step is later followed by the need to {\it read} the stored quantum state. Both protocols~\cite{Julsgaard.PhysRevLett.110.250503,Afzelius.NJP.1367-2630-15-6-065008(2013)} propose to apply sequences of $\pi$ pulses to the spins, combined with dynamical tuning of the resonator frequency~\cite{PalaciosLaloy2008,sandberg_tuning_2008} and quality factor~\cite{Yin.PhysRevLett.110.107001(2013),Wenner.PhysRevLett.112.210501(2014)} in order to trigger the rephasing of the spins, resulting in the emission of an echo at a chosen time that faithfully reproduces the initial quantum state.

\begin{figure}[t]
  \centering
  \includegraphics[width=89mm]{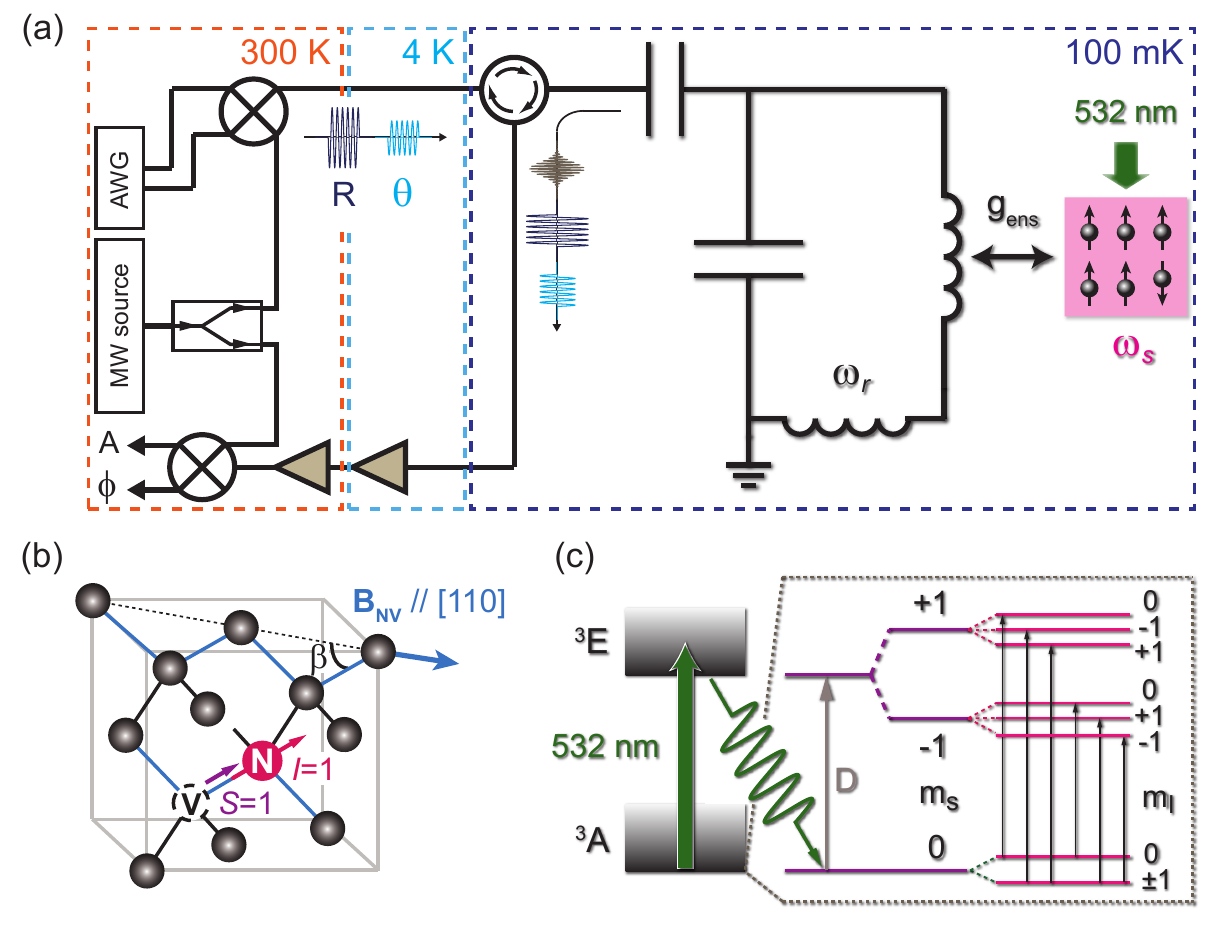}
  \caption{(a) Experimental setup and principle of the experiment. An ensemble of $\sim 10^{10}$ spins is inductively coupled to a planar superconducting LC resonator of frequency $\omega_r$ (with a collective coupling constant $g_{ens}$), cooled at $10$\,mK. The resonator is measured in reflection through an input coupling capacitance. Microwave pulses are produced by mixing a continuous microwave source with dc pulses generated by an arbitrary waveform generator (AWG). They drive the spins via the microwave current induced in the resonator inductance. The reflected microwave signal (including the emitted echo) is amplified at low-temperature and demodulated at room-temperature, yielding its amplitude $A(t)$ and phase $\phi(t)$. (b) The spins are Nitrogen-Vacancy color centers in diamond, which consist of a nitrogen impurity next to a vacancy of the diamond lattice. A dc magnetic field $B_{NV}$ applied parallel to the chip along the $[110]$ crystalline axis so that only NV centers whose axis are non-orthogonal to the field (shown in blue on the figure) are Zeeman-shifted and contribute to the signal. Laser pulses can be sent onto the diamond via a direct optical access to the cryostat mixing chamber. (c) NV centers energy levels in a weak magnetic field. The electronic ground state is a spin triplet $S=1$, with a zero-field splitting $D / 2\pi = 2.88$\,GHz, coupled by hyperfine interaction to the $^{14}N$ nuclear spin triplet $I=1$. This splits each of the $\ket{m_S=0} \rightarrow \ket{m_S = + 1}$ transitions into a triplet of lines.}
		\label{fig1}
	\end{figure}

Whereas the transfer of a qubit state into a spin ensemble has been demonstrated experimentally~\cite{Kubo.PhysRevLett.107.220501(2011),zhu_coherent_2011,SaitoPRL111-107008(2013),Tabuchi.arxiv.1410.3781v1}, implementing the {\it read} step remains the major obstacle to an operational microwave quantum memory. An intermediate goal consists in storing a classical microwave pulse with an ultra-low power corresponding to an average of $1$ photon in the resonator and to retrieve it as an echo after a refocusing pulse, as was achieved at optical frequencies~\cite{Riedmatten.Nature.456.773(2008),Jobez.arxiv.1501.03981v1}. First results in this direction were obtained using ensembles of negatively-charged nitrogen-vacancy (NV) colour centers in diamond~\cite{GruberScience276.2012(1997),Grezes.PHYSREVX4.021049(2014)} and of rare-earth ions in a $\mathrm{Y}_2 \mathrm{Si} \mathrm{O}_5$ crystal~\cite{Probst.Arxiv.1501.01499v2}. The NV's electronic spin is a spin triplet ($S=1$) well suited for a quantum memory because of its long coherence times in pure crystals~\cite{Balasubramian.NatMat.8.383(2009),Bargill.NatCom.4.1743(2013)} and the possibility of repumping it into its ground state $m_S=0$ by optical irradiation at $532$\,nm~\cite{Manson.PhysRevB.74.104303(2006)} (see Fig.~\ref{fig1}). In~\cite{Grezes.PHYSREVX4.021049(2014)}, successive low-power microwave pulses were stored in an NV ensemble, and retrieved later as a series of echoes after a refocusing microwave pulse was applied. A key aspect in this experiment was an active reset of the NVs to increase the repetition rate of successive experimental sequences to obtain sufficient statistics; this was achieved by applying optical pumping laser pulses injected through an optical fiber introduced in the cryostat. The echo efficiency, defined as the ratio of the echo energy and the stored pulse energy, was, however, not sufficient in~\cite{Grezes.PHYSREVX4.021049(2014)} to observe an echo below $100$ photons on average in the resonator.

Here, using a sample with a longer coherence time and an improved optical pumping scheme, we increase the echo efficiency and the storage time by one order of magnitude. This allows us to observe an echo with an initial pulse power corresponding to on average only one photon in the resonator. The diamond single crystal was synthesized by the temperature gradient method at high pressure and high temperature (HPHT) using $99.97\%$ $^{12}$C-enriched pyrolytic carbon prepared from $^{12}$C-enriched methane as a carbon source~\cite{Nakamura.DiamondMaterials.16.1765(2007)}, resulting in a nominal $300$\,ppm concentration of $^{13}C$ nuclei. The original $1.4$\,ppm concentration of substitutional nitrogen impurities (so-called P1 centers) was partially converted into NV centers by $2$\,MeV electron irradiation at room temperature followed by annealing for $2$ hours at $1000^{\circ}$C, yielding a final $[NV^-]=0.4$\,ppm and $[P1] = 0.6$\,ppm. A scheme of the experimental setup is shown in Fig.~\ref{fig1}a. The diamond is glued onto the inductance of a superconducting planar lumped-element LC resonator patterned in a niobium thin-film on a silicon substrate. Microwave pulses can be sent to the resonator input; the reflected signal is amplified at $4$\,K and its amplitude and phase are measured at room-temperature by a homodyne detection setup. For resetting the spins, laser pulses at $532$\,nm can be sent onto the sample through direct optical access in the dilution cryostat.

\begin{figure}[t]
  \centering
  \includegraphics[width=89mm]{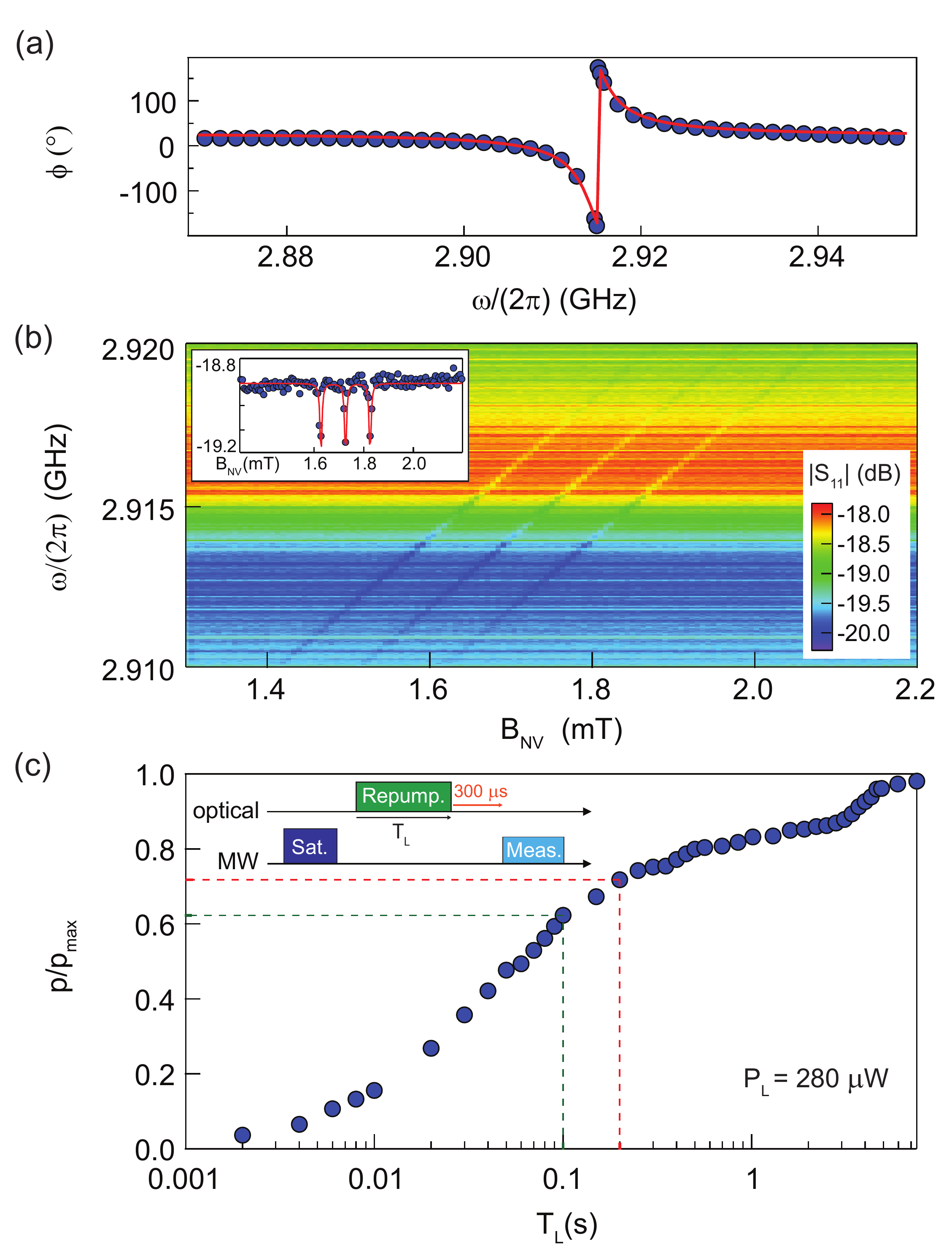}
  \caption{(a) Phase $\phi$ of the resonator reflection coefficient $S_{11}(\omega)$ measured with a vector network analyzer (blue dots), yielding $\omega_r / 2\pi = 2.915$\,GHz and $Q=650$ (red line is a fit to the data). (b) $S_{11}(\omega)$ as a function of $B_{NV}$ around $1.8$\,mT, showing the NV centers as a triplet of absorption dips. The inset shows $|S_{11}|(B_{NV})$ at $\omega/2\pi = 2.915$\,GHz; blue dots are data, and red line is a fit to a sum of three Lorentzians with linewidth $0.012$\,mT. (c) Measured spin polarization for a laser pulse of power $P_L = 280 \mu \mathrm{W}$ as a function of its duration $T_L$, renormalized to its maximal value. Dashed red and black lines indicate the laser pulse durations $T_L=0.2$ and $0.1$\,s used in the experiments shown in Figs.~\ref{fig3} and ~\ref{fig4}, corresponding to relative polarizations $p/p_{max}=0.72$ and $0.62$.}
		\label{fig2}
	\end{figure}

The resonator reflection coefficient $S_{11}(\omega)$, measured with a network analyzer, is shown in Fig.~\ref{fig2}a, yielding the resonance frequency $\omega_r / 2\pi = 2.915$\,GHz and quality factor $Q=650$, fixed by the coupling to the measurement line through the input capacitor. NV centers are detected by their absorption of the microwave whenever their transition frequency matches $\omega_r$. The energy levels of the NV centers are schematically shown in Fig.~\ref{fig1}c. The electronic spin is coupled by hyperfine interaction to the spin-triplet ($I=1$) nuclear spin of the $^{14}N$ atom, resulting in a splitting of the $\ket{m_S = 0} \rightarrow \ket{m_S = + 1}$ transition into three resonances separated by $2.2$\,MHz corresponding to the three different $m_I$ states of the $^{14}N$. A dc magnetic field $B_{NV}$ is applied parallel to the chip, along the $[110]$ crystalline axis of the diamond. Out of the four possible orientations of NV centers along $[111]$ crystalline axes, two are perpendicular to the field and are therefore not Zeeman-shifted, so that they do not contribute to the signal. The remaining two families are brought into resonance with $\omega_r$ at $B_{NV} \sim 1.8$\,mT, as shown in Fig.~\ref{fig2}b where the hyperfine triplet is clearly seen as dips in $|S_{11}(\omega)|$ when they cross the resonance. A $\sim 200$\,kHz Full-Width Half Maximum (FWHM) linewidth is measured for each line of the triplet, much narrower than in previous work~\cite{Kubo.PhysRevLett.107.220501(2011),Grezes.PHYSREVX4.021049(2014)}, due to a lower P1 center concentration and to the isotopic enrichment in $^{12}C$.

For optical pumping of the NV centers, the laser beam is focused to a $0.6$\,mm diameter at the sample level, with a power of $0.28$\,mW. In this geometry, it is straightforward to optimize the laser beam position on the sample in order to minimize the amount of power needed to reset the spins in their ground state. The efficiency of the optical pumping is measured as explained in~\cite{Grezes.PHYSREVX4.021049(2014)}. The experimental sequence includes an initial strong microwave pulse that saturates the spins, followed by a laser pulse of varying duration $T_L$. After a $300\,\mu \mathrm{s}$ delay necessary for relaxation of the quasiparticles generated in the superconducting thin film and in the silicon substrate, the reflected amplitude of a few-photon microwave pulse reveals the spin polarization. The extracted polarization level $p(T_L)$ is shown in Fig.~\ref{fig2}c. It changes only slightly above $T_L=1$\,s, indicating that the maximum NV polarization possible with irradiation at $532$\,nm ($p_{max}=90\%$ according to~\cite{Robledo.Nature.477.574(2011)}) is reached. Compared to earlier work~\cite{Grezes.PHYSREVX4.021049(2014)} where the laser position could not be optimized, the maximum polarization can be reached with $\sim 20$ times lower pulse energy. This makes it possible to perform the experiments at a faster repetition rate ($0.2$\,Hz) and at a cryostat temperature of $100$\,mK instead of $400$\,mK~\cite{Grezes.PHYSREVX4.021049(2014)}.

\begin{figure}[t!]
  \centering
  \includegraphics[width=89mm]{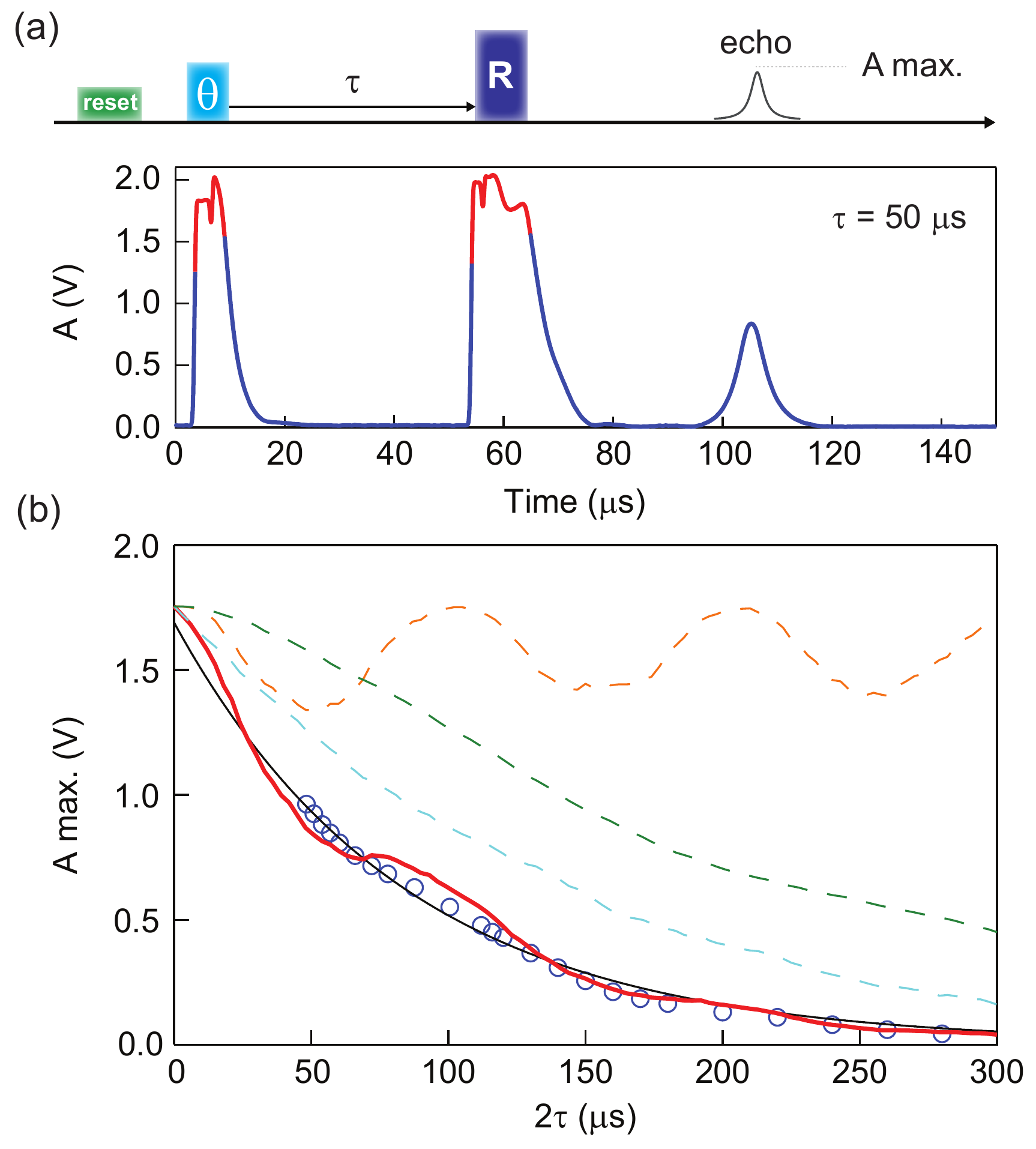}
  \caption{(a) Hahn echo sequence. The spins are first reset in their ground state by a laser pulse of power $280 \mu \mathrm{W}$ and duration $0.2$\,s. A first $1 \mu \mathrm{s}$ microwave pulse $\theta$ at $\omega_r$, of power $-71$\,dBm, induces a transverse magnetization which decays within $T_2^*$. A $1 \mu \mathrm{s}$-long microwave refocusing pulse ($R$) of power $-20$\,dBm is applied at $\tau = 50 \mu \mathrm{s}$, which rephases the spins at $2 \tau$. The microwave amplitude (blue curve) shows both the reflection of the two microwave pulses driving the spins (with their amplitude trimmed by saturation of our detection chain, as indicated in red), as well as the echo emitted at $2\tau$ upon rephasing of the spins. (b) Measured decay of the echo amplitude $A$ as a function of $2 \tau$ (open circles). Calculated decay due to a bath of $213$\,ppm of $^{13}C$ (dashed orange curve), $0.6$\,ppm of $P1$ centers causing spectral diffusion (dashed green line), $0.2$\,ppm of NV centers causing instantaneous diffusion (dashed blue line), and the combination of the three contributions (solid red line). The theory curves have been scaled in amplitude according to the data.  An exponential fit (black solid line) yields a coherence time $T_2 = 84 \mu \mathrm{s}$.}
		\label{fig3}
	\end{figure}

High-power Hahn echoes are measured at $B_{NV}=1.74$\,mT, using microwave pulses at $\omega = \omega_r$ according to the sequence shown in Fig.~\ref{fig3}a. The sequence starts with a laser pulse of duration $T_L = 0.2$\,s, resulting in a spin polarization $p=0.72 \, p_{max} = 0.65$. At $t=0$, a first pulse generates a transverse magnetization in the ensemble, followed by a refocusing microwave pulse at $t=\tau$ which induces rephasing of the spins at $2 \tau$ and emission of a spin-echo into the measuring line, as seen in Fig.~\ref{fig3}a. Note that due to spatial inhomogeneity of the microwave field generated by the planar inductance, it is not possible to apply a well-defined Rabi angle to all the spins, which results in a reduced echo visibility. The echo amplitude is measured as a function of the delay $2\tau$ between the first pulse and the echo, and is found to decay approximately exponentially with a time constant $T_2 =84 \mu \mathrm{s}$ (see Fig.~\ref{fig3}b). Decoherence occurs due to dipolar interactions with the bath of paramagnetic species present in the sample ($^{13}C$ nuclei, P1 centers, and NV centers), whose dynamical evolution causes a randomization of the phase acquired by NV centers during the two halves of the spin-echo sequence. The $^{13}C$ nuclei bath precesses at the Larmor frequency $\gamma_n B_{NV} = 2 \pi \cdot 130$\,kHz ($\gamma_n$ being the $^{13}C$ gyromagnetic ratio), giving rise to a characteristic oscillation pattern~\cite{Maze.PhysRevB.78.094303(2008),Stanwix.PhysRevB.82.201201(2010),Zhao.PhysRevB.85.115303(2012)} in the spin-echo amplitude, visible in the data of Fig.~\ref{fig3}b. The dynamics due to flip-flop events within the P1 center bath is responsible for a decoherence process knwon as spectral diffusion~\cite{deLange01102010}. Finally, the bath consisting of NV centers at frequency $\omega_r$ (only half of the total NV concentration) unavoidably undergoes spin flips due to the refocusing pulse itself, which constitutes an efficient decoherence process called instantaneous diffusion~\cite{tyryshkin_electron_2012}. The various contributions of each bath were calculated using the cluster-correlation expansion method~\cite{Yang.PhysRevB.78.085315(2008),Yang.PhysRevB.79.115320(2009)}, with concentrations $[P1] = 0.6$\,ppm, $[NV^-] = 0.4/2=0.2$\,ppm, and $[^{13}C] = 213$\,ppm, compatible with the sample parameters. Good agreement with the data is obtained (see Fig.~\ref{fig3}b). Overall, dipolar interactions between NV centers appear to be the dominant source of decoherence in our experiment.

\begin{figure}[t!]
  \centering
  \includegraphics[width=89mm]{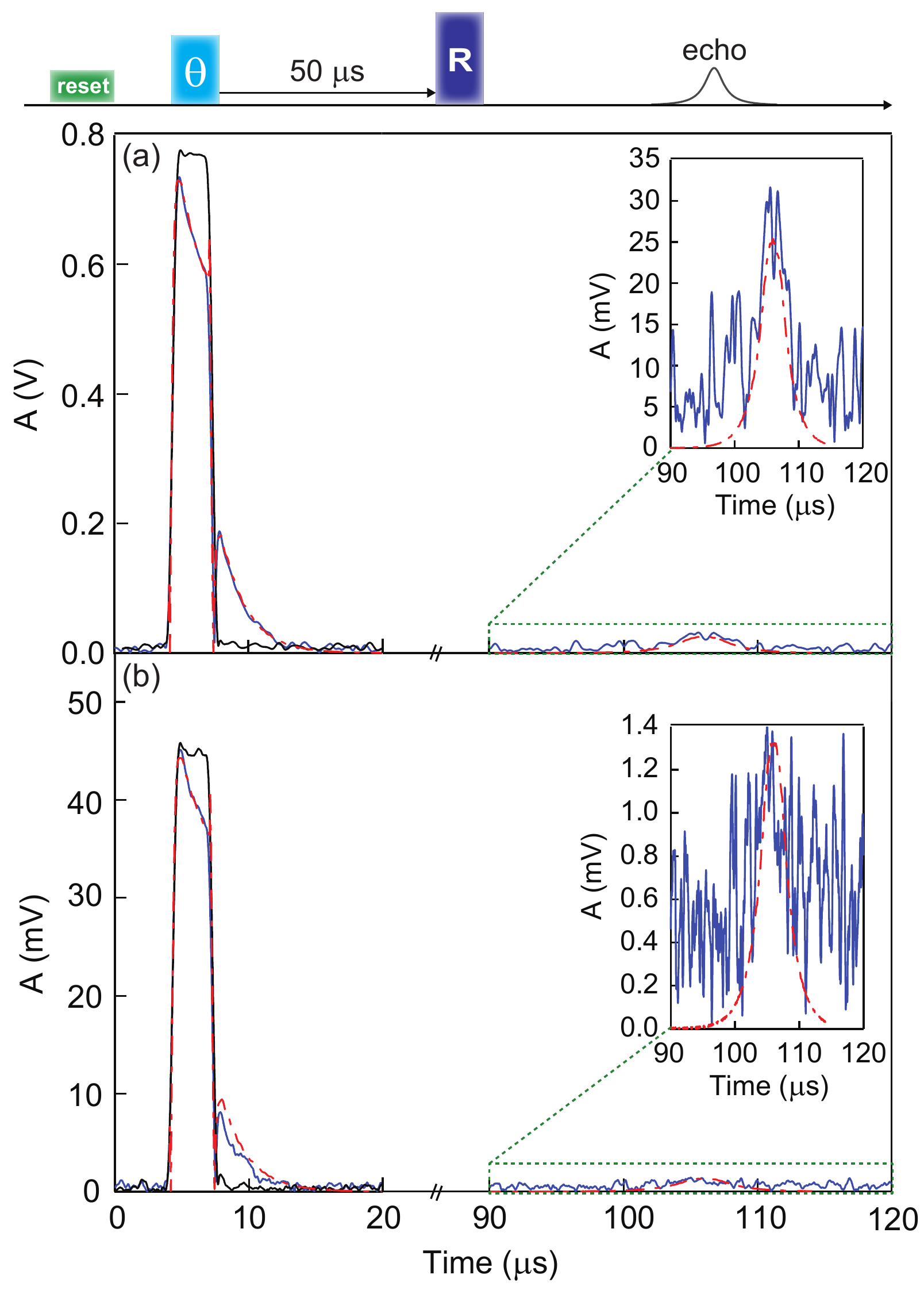}
  \caption{Spin-echo at the few-photon level. The experimental sequence is the same as in Fig.~\ref{fig3}, but with an initial microwave pulse having a power corresponding to (a) $60$ and (b) $1$ photon inside the resonator. The signal was averaged over $3 \times 10^4$ (a) and $5 \times 10^5$ (b) sequences, with a repetition rate of $5$\,Hz (a) and $10$\,Hz (b), limited by the laser pulse duration. Blue solid lines are experimental data, red dashed-dotted lines are the results of simulations as explained in the text.}
		\label{fig4}
	\end{figure}

Since the echo efficiency was limited by the finite spin coherence time in earlier work~\cite{Grezes.PHYSREVX4.021049(2014)}, a significant improvement is expected with this new sample. The echo efficiency is first measured by performing a Hahn echo sequence with a low-power microwave pulse. The experimental sequence, shown in Fig.~\ref{fig4}, starts with a $0.2$\,s repumping laser pulse, followed after $300\, \mu\mathrm{s}$ by a microwave pulse populating the resonator with on average $60$ photons, and, $\tau = 50 \mu\mathrm{s}$ later, by a refocusing pulse. A spin-echo is detected at $t=2\tau$. The efficiency, defined as the energy recovered during the echo divided by the absorbed energy, reaches $E= 0.3 \%$.

The whole spin-echo sequence was simulated by numerical integration of the coupled differential equations obtained after discretizing the resonance frequency and coupling constant distribution in the spin ensemble (see~\cite{Julsgaard.PhysRevLett.110.250503}). A $130$\,kHz linewidth of each HF peak and a collective coupling constant $g_{ens}/2\pi = 410$\,kHz (when all spins are polarized), are compatible with the experimentally determined parameters and yield quantitative agreement with the shape of both the absorbed microwave pulse and the spin-echo amplitude. The discrepancy noted in~\cite{Grezes.PHYSREVX4.021049(2014)} is absent in the present experiment, probably because decoherence is negligible during the driven evolution. The finite $T_2$ and the imperfect $\pi$ pulse due to the spread in Rabi frequencies are the main factors limiting $E$, while the finite cooperativity $C =0.22$~\cite{Julsgaard.PhysRevLett.110.250503} limits both the absorption and the echo emission out of the cavity.

Owing to the larger value of $E$, it becomes possible to reach the level where a spin-echo can be observed for an initial pulse populating the resonator with only a single microwave photon on average. This is shown in Fig.~\ref{fig4}b. Note that a shorter repumping time of $0.1$\,s (with the same laser power) was used, in order to enable a larger number $5 \times 10^5$ of repetitions of the experiment. The shorter repumping step yields a lower spin polarization $p=0.56$ as shown in Fig.~\ref{fig2} and a lower cooperativity of $0.19$, which results in a correspondingly lower echo efficiency than in Fig.~\ref{fig4}a. These results are again quantitatively reproduced by the simulations with the same parameters as mentioned earlier, using the experimentally determined repumping efficiency.

The coherence times demonstrated in this experiment match those requested in a realistic quantum memory protocol~\cite{Julsgaard.PhysRevLett.110.250503}, which suggests that a first implementation is within reach. The remaining challenges are the improvement of the refocusing pulse using adiabatic passage as demonstrated recently~\cite{Sigillito.APL.104.104.22407(2014)}, and the integration of dynamical tuning of the resonator frequency and quality factor with more elaborate spin-echo sequences. The latter is needed in particular to silence the echo emission~\cite{McAuslan.PhysRevA.84.022309(2011),Damon.NewJPhys.13.093031(2011)} in-between the two $\pi$ pulses in the course of the {\it read} step of the procotol. The microwave currents needed to drive the spins during the refocusing pulses are much stronger than typical Josephson junctions critical currents. This precludes the use of integrated SQUIDs as in~\cite{Kubo.PhysRevLett.107.220501(2011),Kubo.PhysRevA.85.012333}, while a combination of coupled linear and tunable resonators~\cite{Grezes.PhD(2014)} may be employed as tuning elements in the resonator.

In conclusion we report the measurement of a spin-echo with an initial microwave pulse at the single-photon level. The figures of merit reached are sufficient to envision first implementations of a spin-ensemble multi-mode quantum memory for superconducting qubits.

\textbf{Acknowledgements} We acknowledge technical support from P. S{\'e}nat, D. Duet, J.-C. Tack, P. Pari, P. Forget, as well as useful discussions within the Quantronics group and with A. Dr{\'e}au, J.-F. Roch, T. Chaneli{\`e}re and J. Morton. We acknowledge support of the French National Research Agency (ANR) with the QINVC project from CHISTERA program, of the European project SCALEQIT, of the C'Nano IdF project QUANTROCRYO, of JST, and of JSPS KAKENHI (no. 26246001). Y. Kubo acknowledges support from the JSPS, and B. Julsgaard and  K. M{\o}lmer from the Villum Foundation.

\end{document}